\documentstyle[12pt]{article}
\newcommand{\lettersection}[1]{\noindent {\it #1} \\}
\newcommand{\be}{\begin{eqnarray}}
\newcommand{\ee}{\end{eqnarray}}
\newcommand{\half}{{\textstyle\frac{1}{2}}}
\newcommand{\textfrac}[2]{{\textstyle\frac{#1}{#2}}}
\newcommand{\Nc}{N_{\rm C}}

\newcommand{\partialslash}{\partial\hspace{-.5em}/\hspace{.15em}}
\newcommand{\pslash}{p\hspace{-.5em}/\hspace{.15em}}
\newcommand{\ie}{{\em i.e.}}
\newcommand{\cf}{{\em cf.}}
\newcommand{\pifac}{\frac{1}{16\pi^2}}
\newcommand{\dfour}[1]{\!\frac{d^4 #1}{(2\pi)^4}}
\newcommand{\ngG}{{\bf\sf G}}
\newcommand{\D}{\Delta}
\newcommand{\Dt}{\tilde{\Delta}}
\newcommand{\lh}[1]{\left(\frac{\lambda^{#1}}{2}\right)}
\newcommand{\sfa}{{\sf a}}
\newcommand{\sfb}{{\sf b}}
\begin{document}
\rightline{UNIT\"U-THEP-4/1993}
\rightline{April 1993}
\vspace{.4in}
\begin{center}
\begin{large}
{\bf Diquark electromagnetic form factors in a Nambu--Jona-Lasinio
model$^\dagger$} \\
\end{large}
\vspace{.8in}
{\bf C. Weiss$^{\rm a, b}$, A. Buck, R. Alkofer and H. Reinhardt} \\
\vspace{0.3in}
Institut f\"ur Theoretische Physik \\
Universit\"at T\"ubingen \\
Auf der Morgenstelle 14 \\
D--7400 T\"ubingen \\
Germany
\end{center}
\vspace{0.4in}
\begin{abstract}
\noindent
Electromagnetic properties of diquarks are investigated in the framework
of a color--octet Nam\-bu--Jona-Lasinio model, which describes baryons as
bound states of diquarks and quarks. We
calculate the electromagnetic form factors of scalar and axial vector
diquark bound states using the gauge--invariant proper--time regularization.
The nucleon charge radii and magnetic moments are estimated in a
simple additive diquark--quark model. This picture reproduces the qualitative
features such as the negative charge radius of the neutron. Within this
model axial vector diquarks are seen to be important.
\end{abstract}
\vfill
\rule{2.in}{.005in}
\\
\noindent
{\footnotesize $^\dagger$ Supported by COSY under contract 41170833
and by the DFG under contract Re 856/2--1} \\
{\footnotesize $^{\rm a}$ Supported by a research fellowship of the Deutsche
Forschungsgemeinschaft (DFG)} \\
{\footnotesize $^{\rm b}$ E-mail: weiss@mailserv.uni-tuebingen.de} \\
\eject
%
%
\lettersection{1.\ Introduction}
A prime objective of hadronic physics is the understanding of
baryons as bound states of QCD. Since the direct application
of QCD in the nonperturbative regime is extremely difficult,
models for baryons have been constructed which preserve certain features
of QCD believed to be essential. Such models include the non-relativistic
quark model, in which baryons are described as constituent quarks
moving in a phenomenological potential, or soliton models based on the
limit $N_{\rm C}\rightarrow\infty$.
In both these approaches the description of baryons is essentially
of a static nature, which entails serious difficulties due to the lack
of proper quantization of the translational motion. A fully relativistic
description of baryons can be accomplished in an approach in which
baryons are considered as bound states of diquarks and
quarks. Evidence for correlated diquark states in baryons is found
in deep--inelastic lepton scattering \cite{landshoff} and in hyperon weak
decays \cite{stech}. Attempts have been made to describe diquarks and
baryons in non-local approximations to QCD \cite{cahill}. A more tractable
approach is based on a local effective quark model, a Nambu--Jona-Lasinio
(NJL--) model with a color--octet current--current interaction. This model
arises as the leading term in a current expansion of the quark--quark
interaction and can also be motivated semiclassically \cite{cairo}. It
describes the spontaneous breaking of chiral symmetry, \ie , the dynamical
generation of quark masses. Furthermore, after Fierz transformation of the
interaction the color-singlet part of the model can be bosonized exactly and
provides a realistic description of meson
properties \cite{ebert,alkofer,lutz}. Diquark bound states have been
considered in \cite{thorsson,vogl}. In order to study baryons, the NJL model
with color-octet interaction has been transformed into an effective theory of
meson and baryon fields by use of functional integral
techniques \cite{reinhardt}. In this theory bound states of diquarks and
quarks occur due to quark exchange. In \cite{buck} the Faddeev--type equation
for the binding of a scalar diquark and a quark has been solved using a static
approximation for the quark exchange interaction. The resulting masses of the
$\half^+$--baryon octet are in satisfactory agreement with
experiment\footnote{Recently, direct numerical solutions to the Faddeev
equation for the nucleon have been obtained \cite{benz}, which support the
static approximation \cite{buck}.}.
\par
Our aim is to extend this approach to the study of the low--energy
electromagnetic structure of baryons. As a first step towards describing
baryon form factors in the effective hadron theory derived from the NJL
model, we calculate here the on-shell electromagnetic form factors of the
constituent diquarks. The relevant diquark bound states occur in the scalar
($0^+$) and axial vector ($1^+$) channel. The diquark form factors are a
crucial ingredient in the baryon form factor and contain information about
the sizes of the correlated diquark states. To ensure gauge invariance, the
NJL quark loop is regularized using Schwinger's proper--time
method \cite{schwinger}. We also
compare with the results obtained with the usual sharp--cutoff
regularization. With the diquark form factors we then estimate the nucleon
charge radii and magnetic moments in an additive model of extended diquarks
and pointlike valence quarks, using the spin--flavor wave functions of the
non-relativistic quark model. Such a picture may be regarded as a crude
approximation to the baryon wave function.
\vspace{0.2cm} \\
\lettersection{2.\ The model}
The basis of our description is an effective quark Lagrangian of the NJL
type,
\be
{\cal L}_q &=& \bar q (i\partialslash - m^0 ) q - \half g j_\mu^A j^{\mu A} ,
\label{njl_quark}
\ee
where $j^{\mu A} = \bar q \gamma^\mu \half\lambda^A q$ is the color-octet
current of the quark field, $g$ an effective coupling constant and $m^0$ the
current quark mass matrix. By way of Fierz transformations in color, flavor
and Dirac space the interaction in eq.(\ref{njl_quark}) can be rewritten
equivalently as an interaction in the color-singlet ($1_{\rm C}$)
quark--antiquark and color-antitriplet ($\bar 3_{\rm C}$) quark--quark
channel \cite{reinhardt},
\be
-\textfrac{3}{2} g j_\mu^A j^{\mu A} &=&
g_1 (\bar q \Lambda_\alpha q)(\bar q \Lambda^\alpha q)
\,\, + \,\, g_2 (\bar q \Gamma_\alpha C \bar q^T )(q^T C\Gamma^\alpha q) .
\label{fierz}
\ee
Here, $C = i\gamma^2\gamma^0$ is the charge conjugation matrix and the
vertices $\Lambda^{\alpha} , \Gamma^{\alpha}$ are matrices in
color, flavor and Dirac space,
\be
\Lambda^{\alpha} = 1_{\rm C} \lh{a}_{\rm F} O^{\sf a}, \hspace{2em}
\Gamma^{\alpha}  = \left(\frac{i\epsilon^A}{\sqrt{2}}\right)_{\rm C}
\lh{a}_{\rm F} O^{\sfa}, \hspace{3em} \alpha = (A, a, \sfa ) ,
\label{generators}
\ee
where
\be
O^{\sfa} &\in&  \left\{ 1,\; i\gamma_5 ,\; \frac{i\gamma^\mu}{\sqrt{2}} ,
\; \frac{i\gamma^\mu\gamma_5}{\sqrt{2}} \right\}
\label{dirac_generators}
\ee
and $(\epsilon^A )_{BC} = \epsilon_{ABC}$ are the antisymmetric generators of
the $\bar 3$--representation of $SU(3)_{\rm C}$. From the Fierz
transformation one obtains $g_1 = g_2 = g$. However, the two terms in
eq.(\ref{fierz}) are independently chirally invariant, so that chiral
symmetry does not preclude different effective coupling constants
in the meson and diquark channel.
\par
By introducing collective meson and baryon fields, the generating functional
defined by the quark lagrangian,
eq.(\ref{njl_quark}), has been rewritten as an effective hadron theory
\cite{reinhardt}. In the course of this
reformulation one also introduces collective fields for diquarks,
which form building blocks for baryons. One obtains an effective action of
meson and diquark fields, $\phi^\alpha$ and $\D^\alpha$, coupling to
a baryon source, $\chi$,
\be
{\cal S} &=&
 - \frac{3}{4g} \int\! d^4 x\, \left( \phi^{\alpha}\phi^{\alpha} +
\D^{*\alpha}\D^{\alpha} \right) - \half i\, {\rm Tr}_\Lambda \log \ngG^{-1}
- \int\! d^4 x\, (\bar\chi\;\; \chi^T )\, \ngG_{B}[\D ]
\left( \begin{array}{c} \chi \\ \bar\chi^T \end{array} \right) .
\label{effective_action}
\ee
Here, $\ngG$ is the quark Green's function in the Nambu--Gorkov
formalism of superconduc\-ti\-vi\-ty\footnote{Following ref. \cite{reinhardt}
we denote $2\times 2$--matrices in the Nambu--Gorkov
formalism by bold sans--serif letters.},
\be
\ngG^{-1} &=&
\left(\begin{array}{rr} G^{-1} & \D C^\dagger \\ -C\Dt & -G^{-1\;T}
\end{array}\right) , \;\;\;\;
G^{-1} = i \partialslash - m_0 - \phi - \frac{3}{g} \int\! d^4 x\,
\chi \bar\chi ,
\label{nambu_gorkov_g}
\ee
and the collective meson and diquark fields couple to the quarks through
the vertices $\phi = \phi^{\alpha}\Lambda^{\alpha}$ and
$\D = \D^{\alpha}\Gamma^{\alpha} , \; \Dt\; =\; \D^{*\alpha}\Gamma^{\alpha}$.
In particular, the vacuum value of the scalar meson field
generates the constituent quark mass, $M$.
The form of the baryon propagator, $\ngG_{\rm B}[\D]$ is given
in \cite{reinhardt}.
After integrating out the diquark fields one obtains from
eq.(\ref{effective_action}) an effective hadron theory, in which diquarks
and quarks interact through quark exchange \cite{reinhardt} and form bound
baryon states \cite{buck}.
\par
The quark loop in eq.(\ref{effective_action}) is defined with a cutoff
procedure, which is an important physical ingredient of this effective
model. A simple regularization scheme is a sharp Euclidean cutoff, applied
to the momentum integrals resulting from the expansion of
${\rm Tr}\;\log\ngG^{-1}$
in meson and diquark fields. This method in general breaks gauge invariance,
when an electromagnetic field is introduced. A gauge invariant regularization
is provided by the proper--time method of Schwinger \cite{schwinger},
which defines the real part of the quark loop, after continuation of $\ngG$
to Euclidean space, as
\be
{\rm Re}\,
{\rm Tr}_\Lambda \log \ngG_E &=& \half\int_{\Lambda^{-2}}^\infty\frac{ds}{s}
\,\, {\rm Tr}\,\exp (-s\,\ngG^\dagger_E\ngG_E ) .
\label{proper_time}
\ee
The imaginary part of the Euclidean quark loop, which generates anomalous
(intrinsic parity--violating)  meson and diquark processes, is finite and
left unregularized \cite{ebert}.
\vspace{0.2cm} \\
\lettersection{3.\ Diquark electromagnetic form factors}
An important ingredient in the electromagnetic structure of baryons
are the electromagnetic form factors of the constituent
diquarks. To describe the electromagnetic interactions of diquarks, we
couple an electromagnetic field to the quark fields of
eq.(\ref{njl_quark}), by way of minimal substitution,
$i\partialslash \rightarrow i\partialslash - Q A_\mu \gamma^\mu $,
where $Q = \half e (\lambda^3 + \frac{1}{\sqrt{3}}\lambda^8 )$
is the quark charge matrix. We then expand the effective action,
eq.(\ref{effective_action}), in the background of the electromagnetic field
to second order in the diquark field. (Here, the baryon source is set to
zero, $\chi = 0$, and the meson field is left at its vacuum value.)
In Minkowski space and in momentum representation, this expansion
is of the form
\be
{\cal S} &=& {\cal S}_0
\; +\; \int\dfour{p}\;
\D^*_{\alpha}(-p) {\cal D}^{-1\,\alpha\beta}(p) \D_{\beta}(p)
\nonumber \\
&+& \int\dfour{p}\;\int\dfour{q}
\D^*_{\alpha}(-p - \half q) \D_\beta (p - \half q ) A_\mu (q)
{\cal F}^{\alpha\beta\mu} (p, q) ,
\label{expansion}
\ee
where ${\cal D}$ is the diquark propagator and ${\cal F}$ the electromagnetic
vertex function. The masses of the diquark bound states are determined as the
zeros of the inverse propagator. On-shell diquark form factors are then
obtained by evaluating the vertex function for appropriate incoming and
outgoing four-momenta $p \pm \half q$ on the diquark mass shell and
normalizing the fields in eq.(\ref{expansion}) to unit residue of the
propagator. Note that the electromagnetic couplings of the diquark fields
come entirely from the quark loop of eq.(\ref{effective_action}). The diquark
propagator and electromagnetic vertex function are diagonal in color and of
the form\footnote{The symbol $g^{\sfa\sfb}$ means the metric tensor if
$\sfa , \sfb$ are Lorentz indices, otherwise $\delta^{\sfa\sfb}$.}
\be
{\cal D}^{-1\,\alpha\beta}(p) &=& \delta^{AB} \sum_{ij}\lh{a}_{ij}\lh{b}_{ji}
\left( -\frac{3}{2g_2}\, g^{\sfa\sfb} + I_{ij}^{\sfa\sfb}(p) \right) ,
\label{propagator} \\
{\cal F}^{\alpha\beta\mu}(p, q) &=&
\delta^{AB} \sum_{ij}\lh{a}_{ij} \lh{b}_{ji}
\left( Q_i J_{ij}^{\sfa\sfb\mu}(p, q) -
Q_j J_{ji}^{\sfb\sfa\mu}(-p, q) \right) .
\label{vertex_function}
\ee
In cutoff regularization, the functions $I^{\sfa\sfb}_{ij}$ and
$J^{\sfa\sfb\mu}_{ij}$ are given by the loop integrals
\be
I_{ij}^{\sfa\sfb}(p) &=& \half i \int_{\Lambda^2}\dfour{k}
{\rm tr}_{\rm Dirac}\, [G_i (k - \half p) O^\sfa G_j (k + \half p) O^\sfb] ,
\label{sharp_cutoff}
\\
J_{ij}^{\sfa\sfb\mu} (p, q) &=& \half i \int_{\Lambda^2}\dfour{k}
{\rm tr}_{\rm Dirac}\, [ G_i (k - \half p) O^\sfa
G_j (k + \half p - \half q ) \gamma^\mu G_j (k + \half p + \half q) O^\sfb ] ,
\nonumber
\ee
where $G_i (p) = (\pslash - M_i)^{-1}$ is the quark propagator. In
proper--time regularization corresponding expressions are derived using
standard techniques for the expansion of the time--ordered exponential,
eq.(\ref{proper_time}) \cite{schwinger}. Only the final expressions will
be given below.
\par
We now evaluate eqs.(\ref{propagator}, \ref{vertex_function}) for
the scalar and the axial vector diquark channel, which are relevant to the
description of baryons. We use throughout the proper--time regularization
and only compare the numerical results with those obtained with cutoff
regularization. Since we shall consider only the nucleon later
we assume the isospin limit, $M_u = M_d = M$.
Let us first consider the scalar $ud$--diquark ($O^{\sfa} = i\gamma_5$).
In proper--time regularization one finds, after continuation to Minkowski
space\footnote{We suppress the flavor indices on $I^{\sfa\sfb}$ and
$J^{\sfa\sfb}$ in the following.},
\be
I_{0^+} (p) &=& p^2 A_0 (p^2 ) + 2 M^2 B,
\label{scalar_polarization}
\ee
with
\be
A_0 (p^2 ) &=& \pifac\, \int_0^1 d\alpha\, \Gamma\left( 0,
\frac{M^2 - \alpha (1 - \alpha ) p^2}{\Lambda^2} \right) , \\
B &=& \pifac \Gamma\left( -1, \frac{M^2}{\Lambda^2} \right) .
\nonumber
\ee
The second term in eq.(\ref{scalar_polarization}) is related to the quark
condensate, $\langle\bar u u\rangle = \langle\bar d d\rangle = 4 \Nc M^3 B$.
The scalar diquark mass, $m_{0^+}$, is shown in table 1 for different values
of the effective
diquark coupling constant, $g_2$. Here, the coupling in the meson channel,
$g_1$, is determined by the constituent quark mass through the gap equation,
the cutoff is fixed by fitting the pion decay constant, and $m^0$ is
determined from the pion mass.
Diquark couplings of $g_2 / g_1 \sim 2$ are required to reproduce the
masses of the spin--$\half$ baryons if quark exchange is included \cite{buck}.
A ratio $g_2 / g_1 \sim 2.5$ is needed in an additive diquark--quark
model, see below. Similar values are also required to obtain sufficiently
bound axial diquarks.
%
%
\begin{table}
\centering
\begin{tabular}{|l|c|c|c|c|c|}
\hline
$g_2 /g_1$ & $m_{0^+} /{\rm MeV}$ & $m_{1^+}/{\rm MeV}$ &
$\langle r^2\rangle_{0^+}^{1/2}/{\rm fm}$ &
$\langle r^2\rangle_{1^+}^{1/2}/{\rm fm}$ & $\mu_{1^+}$ \\
\hline
   1    & 700 & --  &  0.55 & --   &  --   \\
   1.5  & 573 & --  &  0.49 & --   &  --   \\
   1.75 & 511 & 795 &  0.48 & 1.71 & 0.97  \\
   2    & 449 & 787 &  0.48 & 1.08 & 0.95  \\
   2.5  & 318 & 764 &  0.48 & 0.75 & 0.91  \\
   3    & 140 & 741 &  0.48 & 0.65 & 0.88  \\
\hline
\end{tabular}
\caption[]{\it The masses of the scalar $ud$-- and the axial vector
$uu$--diquark,
$m_{0^+}$ and $m_{1^+}$, and their electromagnetic charge radii, for various
values of the effective diquark coupling constant, $g_2$. Also shown
is the magnetic moment of the axial $uu$--diquark, $\mu_{1^+}$, in units of
$2Q_u e/2M$. Proper--time regularization is used; the parameters are
$M = 400\,{\rm MeV}, \Lambda = 630\,{\rm MeV}, m^0 = 17\,{\rm MeV}.$}
\end{table}
\par
To obtain the electromagnetic form factor for an on--shell scalar diquark of
mass $m_{0^+}$, we evaluate the vertex function, eq.(\ref{vertex_function}),
for incoming and outgoing momenta $p\pm\half q$ with
$(p\pm\half q)^2 = m_{0^+}^2$, which entails $p\cdot q = 0$. On the mass
shell, the vertex function is transverse to the photon momentum,
\be
J_{0^+}^\mu (p, q ) &=& 2 F_{0^+} (q^2 )\, p^\mu ,
\label{ff_scalar_diquark}
\ee
and the normalized diquark form factor is defined as
$f_{0^+} (q^2 ) = F_{0^+} (q^2 )/Z_{0^+}$, with
$Z_{0^+} = \partial /\partial p^2\, I_{0^+} |_{ p^2 = m_{0^+}^2}$.
In proper--time regularization one finds
\be
F_{0^+} (q^2 ) &=&
A_0 (m_{0^+}^2 ) + (m_{0^+}^2 - \half q^2) C_1 (q^2 ) + \half q^2 C_2 (q^2 )
\label{f_on-shell}
\ee
where
\be
C_{1,\, 2} (q^2 ) &=& \pifac\, \int_0^1 d\beta\, \int_0^{1 - \beta} d\alpha\,
X_{1,\, 2} \, \frac{\exp (-Y^2/\Lambda^2 )}{Y^2} , \label{y2} \\
X_1 &=& \alpha , \hspace{1.5em} X_2 \; = 1 , \hspace{1.5em}
Y^2 \; =\; M^2 - \alpha (1 - \alpha ) m_{0^+}^2
- \beta (1 - \alpha - \beta ) q^2 \nonumber
\ee
The scalar diquark form factor as a function of the photon momentum, $q^2$,
is shown in fig.\ 1, for a diquark mass of $m_{0^+} = 318\,{\rm MeV}$.
In particular, $f_{0^+} (0) = 1$, \ie , the total charge is conserved
as a consequence of the gauge--invariant regularization. The scalar diquark
r.m.s.\ charge radius, $\langle r^2 \rangle_{0^+}
= -6\, \partial /\partial q^2 f_{0^+}|_{q^2 = 0}$ is given in table 1.
%
%
\begin{figure}
\centering
%
%
\setlength{\unitlength}{0.240900pt}
\ifx\plotpoint\undefined\newsavebox{\plotpoint}\fi
\sbox{\plotpoint}{\rule[-0.175pt]{0.350pt}{0.350pt}}%
\begin{picture}(1500,900)(0,0)
\tenrm
\put(850,158){\rule[-0.175pt]{0.350pt}{151.526pt}}
\put(264,158){\rule[-0.175pt]{4.818pt}{0.350pt}}
\put(242,158){\makebox(0,0)[r]{0.6}}
\put(1416,158){\rule[-0.175pt]{4.818pt}{0.350pt}}
\put(264,248){\rule[-0.175pt]{4.818pt}{0.350pt}}
\put(242,248){\makebox(0,0)[r]{0.8}}
\put(1416,248){\rule[-0.175pt]{4.818pt}{0.350pt}}
\put(264,338){\rule[-0.175pt]{4.818pt}{0.350pt}}
\put(242,338){\makebox(0,0)[r]{1}}
\put(1416,338){\rule[-0.175pt]{4.818pt}{0.350pt}}
\put(264,428){\rule[-0.175pt]{4.818pt}{0.350pt}}
\put(242,428){\makebox(0,0)[r]{1.2}}
\put(1416,428){\rule[-0.175pt]{4.818pt}{0.350pt}}
\put(264,517){\rule[-0.175pt]{4.818pt}{0.350pt}}
\put(242,517){\makebox(0,0)[r]{1.4}}
\put(1416,517){\rule[-0.175pt]{4.818pt}{0.350pt}}
\put(264,607){\rule[-0.175pt]{4.818pt}{0.350pt}}
\put(242,607){\makebox(0,0)[r]{1.6}}
\put(1416,607){\rule[-0.175pt]{4.818pt}{0.350pt}}
\put(264,697){\rule[-0.175pt]{4.818pt}{0.350pt}}
\put(242,697){\makebox(0,0)[r]{1.8}}
\put(1416,697){\rule[-0.175pt]{4.818pt}{0.350pt}}
\put(264,787){\rule[-0.175pt]{4.818pt}{0.350pt}}
\put(242,787){\makebox(0,0)[r]{2}}
\put(1416,787){\rule[-0.175pt]{4.818pt}{0.350pt}}
\put(264,158){\rule[-0.175pt]{0.350pt}{4.818pt}}
\put(264,113){\makebox(0,0){-0.2}}
\put(264,767){\rule[-0.175pt]{0.350pt}{4.818pt}}
\put(411,158){\rule[-0.175pt]{0.350pt}{4.818pt}}
\put(411,113){\makebox(0,0){-0.15}}
\put(411,767){\rule[-0.175pt]{0.350pt}{4.818pt}}
\put(557,158){\rule[-0.175pt]{0.350pt}{4.818pt}}
\put(557,113){\makebox(0,0){-0.1}}
\put(557,767){\rule[-0.175pt]{0.350pt}{4.818pt}}
\put(704,158){\rule[-0.175pt]{0.350pt}{4.818pt}}
\put(704,113){\makebox(0,0){-0.05}}
\put(704,767){\rule[-0.175pt]{0.350pt}{4.818pt}}
\put(850,158){\rule[-0.175pt]{0.350pt}{4.818pt}}
\put(850,113){\makebox(0,0){0}}
\put(850,767){\rule[-0.175pt]{0.350pt}{4.818pt}}
\put(997,158){\rule[-0.175pt]{0.350pt}{4.818pt}}
\put(997,113){\makebox(0,0){0.05}}
\put(997,767){\rule[-0.175pt]{0.350pt}{4.818pt}}
\put(1143,158){\rule[-0.175pt]{0.350pt}{4.818pt}}
\put(1143,113){\makebox(0,0){0.1}}
\put(1143,767){\rule[-0.175pt]{0.350pt}{4.818pt}}
\put(1290,158){\rule[-0.175pt]{0.350pt}{4.818pt}}
\put(1290,113){\makebox(0,0){0.15}}
\put(1290,767){\rule[-0.175pt]{0.350pt}{4.818pt}}
\put(1436,158){\rule[-0.175pt]{0.350pt}{4.818pt}}
\put(1436,113){\makebox(0,0){0.2}}
\put(1436,767){\rule[-0.175pt]{0.350pt}{4.818pt}}
\put(264,158){\rule[-0.175pt]{282.335pt}{0.350pt}}
\put(1436,158){\rule[-0.175pt]{0.350pt}{151.526pt}}
\put(264,787){\rule[-0.175pt]{282.335pt}{0.350pt}}
\put(23,472){\makebox(0,0)[l]{\shortstack{$f_{0^+}, \, f_{1^+}^{\rm e}$}}}
\put(850,68){\makebox(0,0){$q^2 \; ({\rm GeV})^2$}}
\put(264,158){\rule[-0.175pt]{0.350pt}{151.526pt}}
\put(1306,722){\makebox(0,0)[r]{$0^+$}}
\put(1328,722){\rule[-0.175pt]{15.899pt}{0.350pt}}
\put(381,276){\usebox{\plotpoint}}
\put(381,276){\rule[-0.175pt]{2.409pt}{0.350pt}}
\put(391,277){\rule[-0.175pt]{2.168pt}{0.350pt}}
\put(400,278){\rule[-0.175pt]{2.168pt}{0.350pt}}
\put(409,279){\rule[-0.175pt]{2.409pt}{0.350pt}}
\put(419,280){\rule[-0.175pt]{2.168pt}{0.350pt}}
\put(428,281){\rule[-0.175pt]{1.084pt}{0.350pt}}
\put(432,282){\rule[-0.175pt]{1.084pt}{0.350pt}}
\put(437,283){\rule[-0.175pt]{2.409pt}{0.350pt}}
\put(447,284){\rule[-0.175pt]{2.168pt}{0.350pt}}
\put(456,285){\rule[-0.175pt]{2.409pt}{0.350pt}}
\put(466,286){\rule[-0.175pt]{2.168pt}{0.350pt}}
\put(475,287){\rule[-0.175pt]{2.168pt}{0.350pt}}
\put(484,288){\rule[-0.175pt]{2.409pt}{0.350pt}}
\put(494,289){\rule[-0.175pt]{1.084pt}{0.350pt}}
\put(498,290){\rule[-0.175pt]{1.084pt}{0.350pt}}
\put(503,291){\rule[-0.175pt]{2.168pt}{0.350pt}}
\put(512,292){\rule[-0.175pt]{2.409pt}{0.350pt}}
\put(522,293){\rule[-0.175pt]{2.168pt}{0.350pt}}
\put(531,294){\rule[-0.175pt]{2.409pt}{0.350pt}}
\put(541,295){\rule[-0.175pt]{2.168pt}{0.350pt}}
\put(550,296){\rule[-0.175pt]{1.084pt}{0.350pt}}
\put(554,297){\rule[-0.175pt]{1.084pt}{0.350pt}}
\put(559,298){\rule[-0.175pt]{2.409pt}{0.350pt}}
\put(569,299){\rule[-0.175pt]{2.168pt}{0.350pt}}
\put(578,300){\rule[-0.175pt]{2.168pt}{0.350pt}}
\put(587,301){\rule[-0.175pt]{2.409pt}{0.350pt}}
\put(597,302){\rule[-0.175pt]{1.084pt}{0.350pt}}
\put(601,303){\rule[-0.175pt]{1.084pt}{0.350pt}}
\put(606,304){\rule[-0.175pt]{2.409pt}{0.350pt}}
\put(616,305){\rule[-0.175pt]{2.168pt}{0.350pt}}
\put(625,306){\rule[-0.175pt]{2.168pt}{0.350pt}}
\put(634,307){\rule[-0.175pt]{1.204pt}{0.350pt}}
\put(639,308){\rule[-0.175pt]{1.204pt}{0.350pt}}
\put(644,309){\rule[-0.175pt]{2.168pt}{0.350pt}}
\put(653,310){\rule[-0.175pt]{2.168pt}{0.350pt}}
\put(662,311){\rule[-0.175pt]{2.409pt}{0.350pt}}
\put(672,312){\rule[-0.175pt]{1.084pt}{0.350pt}}
\put(676,313){\rule[-0.175pt]{1.084pt}{0.350pt}}
\put(681,314){\rule[-0.175pt]{2.409pt}{0.350pt}}
\put(691,315){\rule[-0.175pt]{2.168pt}{0.350pt}}
\put(700,316){\rule[-0.175pt]{2.168pt}{0.350pt}}
\put(709,317){\rule[-0.175pt]{1.204pt}{0.350pt}}
\put(714,318){\rule[-0.175pt]{1.204pt}{0.350pt}}
\put(719,319){\rule[-0.175pt]{2.168pt}{0.350pt}}
\put(728,320){\rule[-0.175pt]{2.168pt}{0.350pt}}
\put(737,321){\rule[-0.175pt]{1.204pt}{0.350pt}}
\put(742,322){\rule[-0.175pt]{1.204pt}{0.350pt}}
\put(747,323){\rule[-0.175pt]{2.168pt}{0.350pt}}
\put(756,324){\rule[-0.175pt]{2.409pt}{0.350pt}}
\put(766,325){\rule[-0.175pt]{1.084pt}{0.350pt}}
\put(770,326){\rule[-0.175pt]{1.084pt}{0.350pt}}
\put(775,327){\rule[-0.175pt]{2.168pt}{0.350pt}}
\put(784,328){\rule[-0.175pt]{2.409pt}{0.350pt}}
\put(794,329){\rule[-0.175pt]{1.084pt}{0.350pt}}
\put(798,330){\rule[-0.175pt]{1.084pt}{0.350pt}}
\put(803,331){\rule[-0.175pt]{2.168pt}{0.350pt}}
\put(812,332){\rule[-0.175pt]{1.204pt}{0.350pt}}
\put(817,333){\rule[-0.175pt]{1.204pt}{0.350pt}}
\put(822,334){\rule[-0.175pt]{2.168pt}{0.350pt}}
\put(831,335){\rule[-0.175pt]{2.409pt}{0.350pt}}
\put(841,336){\rule[-0.175pt]{1.084pt}{0.350pt}}
\put(845,337){\rule[-0.175pt]{1.084pt}{0.350pt}}
\put(850,338){\rule[-0.175pt]{2.168pt}{0.350pt}}
\put(859,339){\rule[-0.175pt]{1.204pt}{0.350pt}}
\put(864,340){\rule[-0.175pt]{1.204pt}{0.350pt}}
\put(869,341){\rule[-0.175pt]{2.168pt}{0.350pt}}
\put(878,342){\rule[-0.175pt]{2.409pt}{0.350pt}}
\put(888,343){\rule[-0.175pt]{1.084pt}{0.350pt}}
\put(892,344){\rule[-0.175pt]{1.084pt}{0.350pt}}
\put(897,345){\rule[-0.175pt]{2.168pt}{0.350pt}}
\put(906,346){\rule[-0.175pt]{1.204pt}{0.350pt}}
\put(911,347){\rule[-0.175pt]{1.204pt}{0.350pt}}
\put(916,348){\rule[-0.175pt]{2.168pt}{0.350pt}}
\put(925,349){\rule[-0.175pt]{1.084pt}{0.350pt}}
\put(929,350){\rule[-0.175pt]{1.084pt}{0.350pt}}
\put(934,351){\rule[-0.175pt]{2.409pt}{0.350pt}}
\put(944,352){\rule[-0.175pt]{1.084pt}{0.350pt}}
\put(948,353){\rule[-0.175pt]{1.084pt}{0.350pt}}
\put(953,354){\rule[-0.175pt]{2.409pt}{0.350pt}}
\put(963,355){\rule[-0.175pt]{1.084pt}{0.350pt}}
\put(967,356){\rule[-0.175pt]{1.084pt}{0.350pt}}
\put(972,357){\rule[-0.175pt]{2.168pt}{0.350pt}}
\put(981,358){\rule[-0.175pt]{1.204pt}{0.350pt}}
\put(986,359){\rule[-0.175pt]{1.204pt}{0.350pt}}
\put(991,360){\rule[-0.175pt]{2.168pt}{0.350pt}}
\put(1000,361){\rule[-0.175pt]{1.084pt}{0.350pt}}
\put(1004,362){\rule[-0.175pt]{1.084pt}{0.350pt}}
\put(1009,363){\rule[-0.175pt]{2.409pt}{0.350pt}}
\put(1019,364){\rule[-0.175pt]{1.084pt}{0.350pt}}
\put(1023,365){\rule[-0.175pt]{1.084pt}{0.350pt}}
\put(1028,366){\rule[-0.175pt]{1.204pt}{0.350pt}}
\put(1033,367){\rule[-0.175pt]{1.204pt}{0.350pt}}
\put(1038,368){\rule[-0.175pt]{2.168pt}{0.350pt}}
\put(1047,369){\rule[-0.175pt]{1.084pt}{0.350pt}}
\put(1051,370){\rule[-0.175pt]{1.084pt}{0.350pt}}
\put(1056,371){\rule[-0.175pt]{2.409pt}{0.350pt}}
\put(1066,372){\rule[-0.175pt]{1.084pt}{0.350pt}}
\put(1070,373){\rule[-0.175pt]{1.084pt}{0.350pt}}
\put(1075,374){\rule[-0.175pt]{1.084pt}{0.350pt}}
\put(1079,375){\rule[-0.175pt]{1.084pt}{0.350pt}}
\put(1084,376){\rule[-0.175pt]{2.409pt}{0.350pt}}
\put(1094,377){\rule[-0.175pt]{1.084pt}{0.350pt}}
\put(1098,378){\rule[-0.175pt]{1.084pt}{0.350pt}}
\put(1103,379){\rule[-0.175pt]{1.204pt}{0.350pt}}
\put(1108,380){\rule[-0.175pt]{1.204pt}{0.350pt}}
\put(1113,381){\rule[-0.175pt]{2.168pt}{0.350pt}}
\put(1122,382){\rule[-0.175pt]{1.084pt}{0.350pt}}
\put(1126,383){\rule[-0.175pt]{1.084pt}{0.350pt}}
\put(1131,384){\rule[-0.175pt]{1.204pt}{0.350pt}}
\put(1136,385){\rule[-0.175pt]{1.204pt}{0.350pt}}
\put(1141,386){\rule[-0.175pt]{2.168pt}{0.350pt}}
\put(1150,387){\rule[-0.175pt]{1.084pt}{0.350pt}}
\put(1154,388){\rule[-0.175pt]{1.084pt}{0.350pt}}
\put(1159,389){\rule[-0.175pt]{1.204pt}{0.350pt}}
\put(1164,390){\rule[-0.175pt]{1.204pt}{0.350pt}}
\put(1169,391){\rule[-0.175pt]{1.084pt}{0.350pt}}
\put(1173,392){\rule[-0.175pt]{1.084pt}{0.350pt}}
\put(1178,393){\rule[-0.175pt]{2.409pt}{0.350pt}}
\put(1188,394){\rule[-0.175pt]{1.084pt}{0.350pt}}
\put(1192,395){\rule[-0.175pt]{1.084pt}{0.350pt}}
\put(1197,396){\rule[-0.175pt]{1.084pt}{0.350pt}}
\put(1201,397){\rule[-0.175pt]{1.084pt}{0.350pt}}
\put(1206,398){\rule[-0.175pt]{1.204pt}{0.350pt}}
\put(1211,399){\rule[-0.175pt]{1.204pt}{0.350pt}}
\put(1216,400){\rule[-0.175pt]{2.168pt}{0.350pt}}
\put(1225,401){\rule[-0.175pt]{1.084pt}{0.350pt}}
\put(1229,402){\rule[-0.175pt]{1.084pt}{0.350pt}}
\put(1234,403){\rule[-0.175pt]{1.204pt}{0.350pt}}
\put(1239,404){\rule[-0.175pt]{1.204pt}{0.350pt}}
\put(1244,405){\rule[-0.175pt]{1.084pt}{0.350pt}}
\put(1248,406){\rule[-0.175pt]{1.084pt}{0.350pt}}
\put(1253,407){\rule[-0.175pt]{1.204pt}{0.350pt}}
\put(1258,408){\rule[-0.175pt]{1.204pt}{0.350pt}}
\put(1263,409){\rule[-0.175pt]{1.084pt}{0.350pt}}
\put(1267,410){\rule[-0.175pt]{1.084pt}{0.350pt}}
\put(1272,411){\rule[-0.175pt]{1.084pt}{0.350pt}}
\put(1276,412){\rule[-0.175pt]{1.084pt}{0.350pt}}
\put(1281,413){\rule[-0.175pt]{2.409pt}{0.350pt}}
\put(1291,414){\rule[-0.175pt]{1.084pt}{0.350pt}}
\put(1295,415){\rule[-0.175pt]{1.084pt}{0.350pt}}
\put(1300,416){\rule[-0.175pt]{1.084pt}{0.350pt}}
\put(1304,417){\rule[-0.175pt]{1.084pt}{0.350pt}}
\put(1309,418){\rule[-0.175pt]{1.204pt}{0.350pt}}
\put(1314,419){\rule[-0.175pt]{1.204pt}{0.350pt}}
\sbox{\plotpoint}{\rule[-0.250pt]{0.500pt}{0.500pt}}%
\put(1306,677){\makebox(0,0)[r]{$1^+$}}
\put(1328,677){\usebox{\plotpoint}}
\put(1348,677){\usebox{\plotpoint}}
\put(1369,677){\usebox{\plotpoint}}
\put(1390,677){\usebox{\plotpoint}}
\put(1394,677){\usebox{\plotpoint}}
\put(381,217){\usebox{\plotpoint}}
\put(381,217){\usebox{\plotpoint}}
\put(401,220){\usebox{\plotpoint}}
\put(421,224){\usebox{\plotpoint}}
\put(442,228){\usebox{\plotpoint}}
\put(462,232){\usebox{\plotpoint}}
\put(482,236){\usebox{\plotpoint}}
\put(503,241){\usebox{\plotpoint}}
\put(523,245){\usebox{\plotpoint}}
\put(543,249){\usebox{\plotpoint}}
\put(564,254){\usebox{\plotpoint}}
\put(584,259){\usebox{\plotpoint}}
\put(604,263){\usebox{\plotpoint}}
\put(624,268){\usebox{\plotpoint}}
\put(644,274){\usebox{\plotpoint}}
\put(664,279){\usebox{\plotpoint}}
\put(684,284){\usebox{\plotpoint}}
\put(704,290){\usebox{\plotpoint}}
\put(724,295){\usebox{\plotpoint}}
\put(744,302){\usebox{\plotpoint}}
\put(764,308){\usebox{\plotpoint}}
\put(783,314){\usebox{\plotpoint}}
\put(803,321){\usebox{\plotpoint}}
\put(822,328){\usebox{\plotpoint}}
\put(842,334){\usebox{\plotpoint}}
\put(862,342){\usebox{\plotpoint}}
\put(881,349){\usebox{\plotpoint}}
\put(900,357){\usebox{\plotpoint}}
\put(919,365){\usebox{\plotpoint}}
\put(938,373){\usebox{\plotpoint}}
\put(957,382){\usebox{\plotpoint}}
\put(976,391){\usebox{\plotpoint}}
\put(994,401){\usebox{\plotpoint}}
\put(1012,410){\usebox{\plotpoint}}
\put(1031,420){\usebox{\plotpoint}}
\put(1049,431){\usebox{\plotpoint}}
\put(1067,441){\usebox{\plotpoint}}
\put(1084,453){\usebox{\plotpoint}}
\put(1101,464){\usebox{\plotpoint}}
\put(1118,477){\usebox{\plotpoint}}
\put(1134,489){\usebox{\plotpoint}}
\put(1150,502){\usebox{\plotpoint}}
\put(1166,516){\usebox{\plotpoint}}
\put(1181,530){\usebox{\plotpoint}}
\put(1196,544){\usebox{\plotpoint}}
\put(1210,559){\usebox{\plotpoint}}
\put(1224,575){\usebox{\plotpoint}}
\put(1237,591){\usebox{\plotpoint}}
\put(1251,607){\usebox{\plotpoint}}
\put(1263,623){\usebox{\plotpoint}}
\put(1274,641){\usebox{\plotpoint}}
\put(1285,658){\usebox{\plotpoint}}
\put(1296,676){\usebox{\plotpoint}}
\put(1306,695){\usebox{\plotpoint}}
\put(1316,713){\usebox{\plotpoint}}
\put(1319,719){\usebox{\plotpoint}}
\end{picture}
\caption{\it The electromagnetic charge form factor of the scalar
$ud$--diquark, $f_{0^+}(q^2 )$, (solid line) and the axial
$uu$--diquark $f_{1^+}^{\rm e}(q^2 )$ (dashed line), for
$m_{0^+} = 318\, {\rm MeV}$ and $m_{1^+} = 764\, {\rm MeV}$
($g_2 / g_1 = 2.5,\,\, M = 400\, {\rm MeV}$). Shown are the results for
proper--time regularization. The curve for the pion form factor would
nearly coincide with the one for the scalar diquark.}
\end{figure}
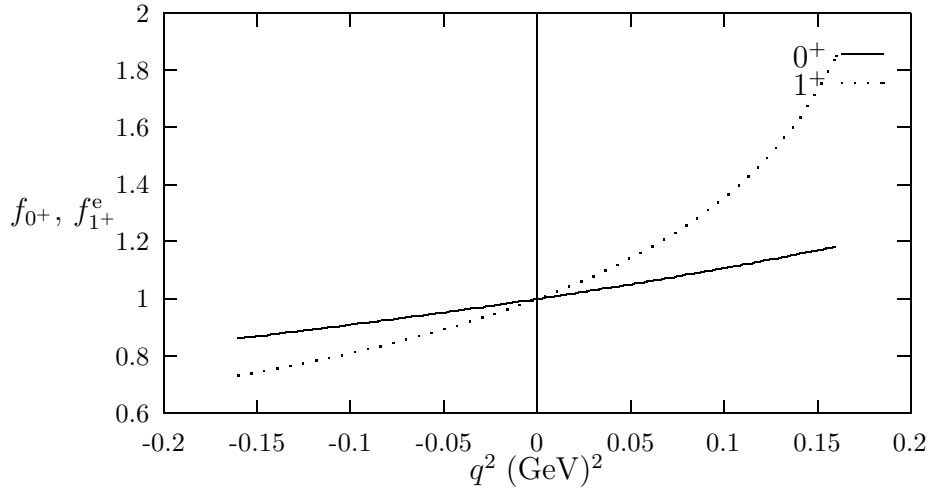
\par
In sharp cutoff regularization the scalar diquark mass and
form factor are evaluated by continuing the integrals (\ref{sharp_cutoff})
to Euclidean space, introducing Feynman parameters and applying the
cutoff after a shift of the integration variable.
Due to the breaking of gauge invariance by the cutoff,
$f_{0^+} (0) \neq 1$ if $m_{0^+}^2 > 0$, so that we have divided
$f_{0^+} (q^2)$ by $f_{0^+} (0)$
to obtain meaningful results. Determining the parameters as above
in the meson sector, somewhat larger ratios $g_2/g_1$ are required
to obtain the same diquark masses as in proper--time regularization.
Nevertheless, comparing the charge radii for scalar diquarks
of the same mass, the results are quite close in both regularization
schemes, \cf\ table 2. The virtue of the gauge--invariant proper--time
regularization, eq.(\ref{proper_time}), is that it is applied at the
level of the action, rather than at the level of individual loop integrals,
which is necessary for the electromagnetic Ward identities to be satisfied.
In particular, with this method the electromagnetic vertices are also
defined off the diquark mass shell.
%
%
\begin{table}
\begin{tabular}{|c||c|c|c||c|c|c|}
\hline
   & \multicolumn{3}{|c||}{\rm proper--time}
   & \multicolumn{3}{|c| }{\rm sharp cutoff} \\
\hline
$m_{0^+} /{\rm MeV}$ &
$m_{1^+}/{\rm MeV}$ &
$\langle r^2\rangle_{0^+}^{1/2}/{\rm fm}$ &
$\langle r^2\rangle_{1^+}^{1/2}/{\rm fm}$ &
$m_{1^+}/{\rm MeV}$ &
$\langle r^2\rangle_{0^+}^{1/2}/{\rm fm}$ &
$\langle r^2\rangle_{1^+}^{1/2}/{\rm fm}$ \\
\hline
400 & 778 & .48 & .90 & 785 & .43 & 1.02 \\
300 & 761 & .48 & .74 & 770 & .42 & 0.79 \\
\hline
\end{tabular}
\caption[]{\it Comparison of the diquark charge radii for proper--time
and cutoff regularization. Here, the ratio $g_2/g_1$ is chosen independently
in both schemes in order to obtain the given scalar diquark masses.}
\end{table}
\par
It is instructive to compare the form factor of the scalar $ud$--diquark
to that of the charged pion \cite{blin}. The only difference between the
diquark and the meson vertex function is the sign of the quark charges in
eq.(\ref{vertex_function}) and a factor $\Nc$ multiplying the
quark loop. However, since the same factor $\Nc$ occurs also in the
meson propagator, this factor cancels if we consider the normalized
on--shell pion form factor. Thus, in the isospin limit the pion form factor
is given by the expression for the diquark form factor, eq.(\ref{f_on-shell}),
evaluated with $m_{0^+}^2 \rightarrow m_\pi^2$. The curve for the pion form
factor as a function of $q^2$ is almost identical to that of the scalar
diquark shown in fig.\ 1.
\par
One may consider the scalar diquark (or pion)
r.m.s.\ charge radius in table 1 as a function of the bound
state mass. For strongly bound diquarks the charge radius is essentially
independent of the diquark mass and thus practically identical to that of
the pion. It is therefore, to good approximation, given
by the result of a gradient expansion around $p^2 = 0$ \cite{volkov},
which in proper--time regularization reads
\be
\langle r^2 \rangle_{0^+} &=& \frac{\Nc}{4\pi^2 f_{\pi}^2}
\exp (-M^2/\Lambda^2 ) .
\ee
In the other extreme, near $m = 2M$, where the diquark would become unbound,
the charge radius
grows like $\langle r^2\rangle_{0^+} \sim (m - 2M)^{-1}$. This is the
behaviour expected for a weakly bound state below a continuum
threshold \cite{jaffe}. Such divergent behaviour would be absent in a
model which incorporates quark confinement.
\par
Let us now consider the axial vector diquark
($O^{\sfa , \sfb} = i\gamma^{\rho , \sigma}/\sqrt{2}$).
With the gauge--invariant proper--time
regularization and in the absence of flavor symmetry breaking, the axial
diquark polarization operator deriving from the quark loop is transverse,
\be
I^{\rho\sigma}_{1^+} (p) &=& p^2 A_1 (p^2 )
(g^{\rho\sigma} - \frac{p^\rho p^\sigma}{p^2})
\label{axial_propagator} ,
\ee
with
\be
A_1 (p^2 ) &=& \pifac\, \int_0^1 d\alpha\, 2 \alpha (1 - \alpha )\,
\Gamma\left( 0, \frac{M^2 - \alpha (1 - \alpha ) p^2}{\Lambda^2}
\right) .
\ee
The longitudinal part of eq.(\ref{propagator}) is thus given entirely by
$(3/2 g_2) g^{\rho\sigma}$. Axial diquark masses are given in table 1 for
various values of
$g_2$. In contrast to the scalar diquark the axial diquark is only weakly
bound for reasonable values of $g_2$. The electromagnetic vertex function,
eq.(\ref{vertex_function}), for on--shell
axial vector diquarks of incoming and outgoing four-momenta $p \pm \half q$,
with $(p \pm \half q)^2 = m_{1^+}^2$, is of the form \cite{lee}
\be
J^{\rho\sigma\mu}_{1^+} (p, q)
&=& 2 g^{\rho\sigma} p^\mu F_{1^+}^{\,\rm e} (q^2 )
- [g^{\rho\mu} (p +\half q)^\sigma + g^{\sigma\mu} (p - \half q)^\rho ]
F_{1^+}^{\rm\, m} (q^2 ). \label{vector_ff}
\ee
In proper--time regularization one finds\footnote{In eq.(\ref{vector_ff}) we
have omitted a term $\propto (p - \half q)^\rho (p + \half q)^\sigma p^\mu$,
since we are interested only in the diquark charge and magnetic moment
form factors.}
\be
F_{1^+}^{\rm e} (q^2 ) &=& \half F_{0^+} (q^2 ) - H_1 (q^2 ) , \\
F_{1^+}^{\rm m} (q^2 ) &=& \half A_0 (m_{1^+}^2 ) - \half A_0 (q^2 )
- \textfrac{1}{4} q^2 C_1 (q^2 ) + \half m_{1^+}^2 C_2 (q^2 ) - H_2 (q^2 ) ,
\ee
where
\be
H_{1,\,  2} &=& \pifac\,\int_0^1 d\alpha
\int_0^{1-\alpha}d\beta\, W_{1, 2}\,
\Gamma\left( 0, \frac{Y^2}{\Lambda^2} \right), \hspace{3em} W_1\, =\,\alpha ,
\hspace{1.5em} W_2\, =\, 2\beta ,
\ee
and $Y^2$ and $F_{0^+} (q^2 )$ are as defined eqs.(\ref{y2}, \ref{f_on-shell})
and evaluated with $m_{0^+} \rightarrow  m_{1^+}$. The normalized
electric and magnetic form factors of the axial vector diquark are obtained
as $f_{1^+}^{\rm e, m} (q^2 ) = F_{\rm e, m}(q^2 )/Z_{1^+}$, where
$Z_{1^+} = \partial /\partial p^2\, (p^2 A_1 ) |_{p^2 = m_{1^+}^2}$.
In particular,
$f_{1^+}^{\,\rm e} (0) = 1$ and $f_{1^+}^{\,\rm m} (0)$ is the magnetic moment
of the $ij$--axial vector diquark in units of $(Q_i + Q_j ) e/2m_{1^+}$
($i, j = u, d$). The electric
form factor describes the charge distribution in the axial diquark and
is shown in fig.\ 1. The charge radius of the axial diquark is larger than
that of the scalar diquark, see table 1. This fact is mainly due to the weaker
binding of the axial diquark. However, even strongly bound axial diquarks
would be larger than scalar diquarks of the same mass. In gradient expansion
at $m_{0^+} = m_{1^+} = 0$ one would find
$\langle r^2 \rangle_{1^+} = \textfrac{27}{20}\langle r^2 \rangle_{0^+}$.
Larger charge radii for the axial diquarks are also found in cutoff
regularization, \cf\ table 2.
The axial diquark magnetic moment is shown in table 1, in units of the
sum of the constituent quark magnetic moments, $(Q_i + Q_j) e/2M$.
The diquark magnetic moment is seen to be lowered compared to the
non-relativistic value as a consequence of the diquark binding.
\vspace{0.2cm} \\
\lettersection{4.\ Estimating the baryon form factor}
Given the diquark electromagnetic form factors we now wish to describe
the electromagnetic interactions of baryons as bound states of constituent
diquarks and quarks. To calculate the baryon electromagnetic form factor in
the effective hadron theory \cite{reinhardt}
based on the NJL model, one would extract from
eq.(\ref{effective_action}) the baryon electromagnetic vertex, which
describes the coupling of the photon to the diquarks, the quarks and to the
quark exchange. One would then take the matrix element of this vertex
between on--shell baryon states, \ie , solutions of the baryon Faddeev
equation \cite{buck}. Since a full calculation of the form factor using bound
state wave functions is rather involved, it is worthwhile to first explore a
simpler picture. Our intention here is to see to what
extent the nucleon charge radii and magnetic moments can be described in an
additive diquark--quark model \cite{vogl}. This approach has
been successful in the description of baryon masses. It is based on the
spin--flavor wave functions of the non-relativistic quark model, in which
two quarks are coupled to a scalar or axial diquark. The diquarks are then
regarded as on-shell constituents, whose intrinsic properties such as
a mass, a charge radius or a magnetic moment, are derived from the NJL
model. Note that as a consequence of the Pauli principle the scalar
diquark is antisymmetric in flavor ($ud$) and has isospin 0, while the axial
diquarks are symmetric ($uu, ud, dd$) and have isospin 1.
\par
We first consider the nucleon electromagnetic charge radii.
In a simple approximation one may neglect the orbital motion of diquarks
and quarks, \ie , take the orbital wave functions of diquarks and quarks
in the baryon Breit frame to be $\delta$--functions centered at the baryon
center--of--mass. The charge density inside the baryon is then
the sum of the extended diquark charge distributions and the pointlike
valence quark charge densities, which do not contribute to the charge radius.
{}From the quark model wave functions one then obtains the
relations\footnote{Here, the isospin limit is assumed, so that the
axial $uu, ud$-- and $dd$--diquarks all have the same charge radius,
$\langle r^2 \rangle_{1^+}$.}
\be
\langle r^2 \rangle_p &=&
\half ( \textfrac{1}{3}\langle r^2 \rangle_{0^+}
+ \langle r^2 \rangle_{1^+} ) , \hspace{3em}
\langle r^2 \rangle_n \; =\;
\half ( \textfrac{1}{3}\langle r^2 \rangle_{0^+}
- \textfrac{1}{3}\langle r^2 \rangle_{1^+} ) . \label{r2_p}
\ee
Note that the diquark charge radii have been defined as derivatives of the
electric form factor normalized to $1$ at $q^2 = 0$, so that the diquark
charges are absorbed in the numerical factors in eq.(\ref{r2_p}).
We now choose the diquark coupling, $g_2$, such as to fit the proton mass,
using the formula of the diquark--quark model,
$m_{p, n} = \half (m_{0^+} + m_{1^+} ) + M$ \cite{vogl}. From table 1
we see
that this requires a value of $g_2 / g_1 = 2.5$. With
the diquark charge radii from table 1 eq.(\ref{r2_p}) gives
$\langle r^2 \rangle_p = (0.57\, {\rm fm})^2$
and $\langle r^2 \rangle_n = - 0.06\, {\rm fm}^2$, which is in
qualitative agreement with the experimental values of
$\langle r^2 \rangle_{p, {\rm exp}}  = (0.85\, {\rm fm})^2$ and
$\langle r^2 \rangle_{n, {\rm exp}}  = -0.11\, {\rm fm}^2$.
This simple picture naturally explains the negative
charge radius of the neutron by the fact that the axial diquark
charge radius is larger than the scalar one. Note that this qualitative
result is independent of the diquark masses.
\par
If one included only scalar diquarks in the above estimate, one would obtain
instead of eq.(\ref{r2_p}) $\langle r^2 \rangle_p = \langle r^2 \rangle_n =
\textfrac{1}{3}\langle r^2 \rangle_{0^+}$, \ie , the neutron charge
radius would come out equal to the proton one. This underscores the importance
of the axial vector diquarks in the additive diquark--quark model. We
remark, however, that in a model with only scalar diquarks a negative charge
radius for the neutron can be obtained if the orbital motion of diquarks
and valence quarks is taken into account, as has been demonstrated in the
framework of the non-relativistic quark model \cite{walle}. The question
of the relative importance of the intrinsic size of the axial diquarks versus
the effects
of the diquark--quark orbital motion can only be answered a posteriori from
the full relativistic bound state wave function containing both scalar and
axial diquark components. Nevertheless, it is encouraging that a simple
additive picture with axial diquark constituents of a size close to that
of the proton is capable of reproducing
both the nucleon mass and the charge radii reasonably well. Moreover, in the
non-relativistic quark model it can be seen that the estimate of the proton
charge radius in
eq.(\ref{r2_p}) is increased by the diquark--quark relative motion, while the
neutron charge radius remains essentially unchanged unless $SU(6)$--breaking
in the spin--flavor wave function is taken into account \cite{walle}.
This strongly suggests that the estimates of the additive model will
be improved by a full calculation using Faddeev wave functions.
\par
Finally, we want to discuss the nucleon magnetic moments in the
diquark--quark picture. Since magnetic moments are described very well
by the non-relativistic quark model, it is important to see how far the
predictions of the diquark--quark picture differ from those of the
quark model. Rewriting the quark model spin--flavor wave functions in terms
of diquarks and quarks one obtains
\be
\mu_p &=& \frac{e}{2M} \left( \textfrac{1}{3} + \textfrac{1}{3} \mu_{1^+}
+ \textfrac{1}{3} \mu_{0^+ - 1^+} \right) , \hspace{2em}
\mu_n\; =\; \frac{e}{2M}\left( -\textfrac{2}{9} - \textfrac{1}{9} \mu_{1^+}
- \textfrac{1}{3} \mu_{0^+ - 1^+} \right) . \label{mag_mom}
\ee
The first term here is the valence quark contribution. Furthermore,
$\mu_{1^+}$ is the magnetic moment of the $(ij)$--axial diquark $(i,j = u, d)$
in units of $(Q_i + Q_j) e/2M$, and $\mu_{0^+ - 1^+}$ is the transition moment
from the $ud$--scalar to the $ud$--axial diquark in units of
$(Q_u - Q_d) e/2M$,
\be
\langle (ud)_{0^+}, S_z = 0 | \mu_{1z} + \mu_{2z}| (ud)_{1^+}, S_z = 0 \rangle
&=& \mu_{0^+ - 1^+} (Q_u - Q_d) (e/2 M) . \label{mut_nr}
\ee
Here, $\mu_{1z} , \mu_{2z}$ are the z--components of the quark magnetic
moment operators. In the case of $SU(6)$ symmetry one has
$\mu_{1^+} = \mu_{0^+ - 1^+} = 1$, which leads to the well--known results
$\mu_p = e/2M,\; \mu_n = -\frac{2}{3} e/2M$.
We now consider eq.(\ref{mag_mom}) in the sense of the additive
diquark--quark model outlined above. The magnetic moment of the axial diquark,
$\mu_{1^+}$, has been calculated above and is given in table 1.
Transitions between scalar and axial vector diquarks due to the
electromagnetic field
are also found with the effective diquark action derived from
the NJL model, eq.(\ref{effective_action}). This is an anomalous
(intrinsic parity--violating) process, which is generated by the
imaginary part of the quark loop; its meson analogue is the process
$\rho\rightarrow\pi\gamma$. Since the definition of $\mu_{0^+ - 1^+}$ in
eq.(\ref{mut_nr}) involves vanishing photon momentum, it is not possible
to describe $\mu_{0^+ - 1^+}$ as a transition between on--shell scalar and
axial
diquarks of unequal mass. To obtain a rough estimate one may evaluate this
vertex for a scalar diquark mass equal to that of the axial diquark. In this
case the transition part of the vertex function, eq.(\ref{vertex_function}),
is given by
\be
J^{\rho\mu}_{0^+ - 1^+} (p, q)
&=& \frac{1}{\sqrt{2}}\,\varepsilon^{\rho\mu\kappa\lambda}
p^\kappa q^\lambda M^{-1} F_{0^+ - 1^+} (p^2 ) ,
\label{f_t}\\
F_{0^+ - 1^+} (p^2 ) &=& \int_0^1 d\alpha
\frac{2 M^2}{M^2 - \alpha (1 - \alpha ) p^2} ,
\nonumber
\ee
and the transition moment corresponds to
$\mu_{0^+ - 1^+} = F_{0^+ - 1^+} (m_{1^+}^2 )/\sqrt{2 Z_{0^+} Z_{1^+}}$. For
the axial diquark masses considered in table 1 $\mu_{0^+ - 1^+}$ is close to
its non-relativistic value of 1; $\mu_{0^+ - 1^+} = 0.97$ for
$g_2 /g_1 = 2.5$. For this value of $g_2 /g_1 = 2.5$ we obtain from
eq.(\ref{mag_mom}) $\mu_p = 0.96\, e/2M, \; \mu_n = -0.65\, e/2M$.
Note that the constituent quark masses used in the NJL model,
$M \sim 400\,{\rm MeV}$, are larger than those of the
quark model, $M = \textfrac{1}{3} m_p \sim 300\,{\rm MeV}$, so that
$\mu_{p, n}$ come out roughly $3/4$ smaller than the quark model prediction.
We emphasize, however, that a proper treatment of the transition contribution
would require the full
baryon wave function with off-shell diquarks. Note that if one included only
scalar diquarks in the additive diquark--quark model, the magnetic moments
would be $\mu_p = \textfrac{2}{3}\, e/2M, \; \mu_n = -\frac{1}{3}\, e/2M$.
The same result would be found if one included axial diquarks but neglected
axial--scalar transitions. These estimates again show the importance of axial
diquarks in the additive model.
\vspace{0.2cm} \\
\lettersection{5.\ Conclusion}
We have calculated the electromagnetic form factors of scalar and axial
vector diquark bound states in a color--octet NJL model. For realistic
diquark masses the scalar diquark charge radius is close to that of the pion.
The axial diquark is weakly bound; its size is of the order of that of the
proton. The nucleon charge radii can be described qualitatively in an additive
diquark--quark model, which may be viewed as a crude approximation to the
baryon wave function. The inclusion of axial vector diquarks in this
scheme is seen to be crucial. For the nucleon magnetic moments scalar--axial
diquark transitions have to be taken into account at least
approximately. It remains to be seen to what extent the results of the
diquark--quark model are improved by a dynamical calculation using the
relativistic baryon wave functions from the Faddeev equation.
\end{document}